\documentstyle[galley,grl,epsfig,amsmath,amssymb]{agu2001}
\begin{document}
\lefthead{Hnat et.~al}
\righthead{Scaling of AE indices}
\title{Scaling in long term data sets of geomagnetic indices and solar wind $\epsilon$ as seen by WIND spacecraft.}
\author{B. Hnat\altaffilmark{1}, S.C. Chapman \altaffilmark{1},
G. Rowlands \altaffilmark{1}, N.W. Watkins \altaffilmark{2} and
M.P. Freeman \altaffilmark{2}}
\altaffiltext{1}{Space and Astrophysics Group, University of Warwick Coventry, CV4 7AJ, UK}
\altaffiltext{2}{British Antarctic Survey, Natural Environment Research Council, Cambridge, CB3 0ET, UK}

\begin{abstract}
We study scaling in fluctuations of the geomagnetic indices ($AE$, $AU$,
and $AL$) that provide a measure of magnetospheric activity and of the
$\epsilon$ parameter which is a measure of the solar wind driver.
Generalized structure function (GSF) analysis shows that fluctuations
exhibit self-similar scaling up to about $1$ hour for the $AU$ index and about
$2$ hours for $AL$, $AE$ and $\epsilon$ when the most extreme fluctuations
over $10$ standard deviations are excluded. The scaling exponents of the GSF
are found to be similar for the three $AE$ indices, and to differ significantly
from that of $\epsilon$. This is corroborated by direct comparison of their
rescaled probability density functions.
\end{abstract}

\begin{article}
\section{Introduction}
The statistical properties of fluctuations in geomagnetic indices and their
relation to those in the solar wind, is a topic of considerable interest
(see, e.g., \citep{sitnov,tsurutani,ukhorskiy,voros}). Scaling has
been identified as a key property of magnetospheric energy release in the form
of bursty bulk flows in the magnetotail \citep{angelopoulos}, ``blobs" in the
aurora \citep{lui}, non-Gaussian fluctuations in geomagnetic indices
\citep{hnat02,hnat03a,consolini96} and in single station magnetometer
data \citep{kovacs,voros}. Models include Self-Organized Criticality (SOC)
\citep{chang03} (see also the review \citep{chapman01}) and multi-fractal
models \citep{kovacs} related to those of turbulence\citep{consolini96,voros}.

These measures of scaling and non-Gaussian fluctuations in magnetospheric
output need to be understood in the context of the system's driver, the solar
wind, which is turbulent and thus also scaling. Other work has focussed on
comparing properties of input parameters such as $\epsilon$ and the indices
($AE$, $AU$ and $AL$) to establish whether they are directly related. However,
these studies have not provided a consistent answer. While \cite{freeman} found
that both the $\epsilon$ and the $AU$ and $AL$ indices exhibited nearly
identical scaling of burst lifetime probability density functions (PDFs),
\cite{uritsky} obtained quite different scalings for $AE$ and the solar wind
quantity $v_xB_{yz}$ using spreading exponent methods motivated by SOC.
\cite{hnat02,hnat03a} used a PDF rescaling technique to characterize the
fluctuation PDF of $4$ years $\epsilon$ data from WIND and a $1$ year data set
of $AE$ indices with fluctuations over a few standard deviations. Direct
comparison of the PDF's functional form suggested close similarity to within
statistical error.

In this paper we use a larger $10$-year data set for the $AE$ indices to obtain
a more accurate statistical determination of the functional form of the PDF of
fluctuations over a more extensive dynamic range, including characterization
of extremal events up to $10$ standard deviations for the first time.
We apply structure functions to characterize and compare both the low and
higher order moments for all quantities. A $4$-year subset of the index data,
corresponding to the same period in the solar cycle as that used to produce
$\epsilon$, is used to facilitate this comparison.
We then verify these results  by direct examination of the fluctuation PDF
using the full $10$-year $AE$ indices dataset.
\begin{figure}
\epsfsize=0.45\textwidth \centerline{
\leavevmode\epsffile{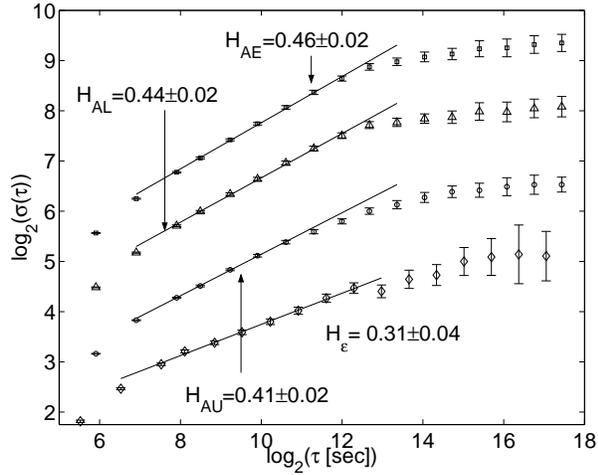}} \caption{Scaling of the standard deviation
of the PDFs of: $\diamond$-$\epsilon$, $\circ$-$AU$ index, $\triangle$-$AL$
index and $\Box$-the $AE$ index. The plots have been offset vertically for
clarity. Error bars are estimated assuming Gaussian statistics for the binned
data.}
\label{fig1}
\end{figure}

\section{Data Sets}
The $AL$, $AU$ and $AE$ index data sets investigated here comprise over $5.5$
million, $1$ minute averaged samples from January $1978$ to December $1988$
inclusive. The $\epsilon$ data set is identical to that used in
\cite{hnat02,hnat03a} and extends from January $1995$ to December $1998$
inclusive. It includes intervals of slow and fast speed streams. $\epsilon$
is defined (see \citep{hnat02}) in SI units as
$\epsilon=v(B^2/\mu_0) l_0^2 \sin^4(\Theta /2)$, where $l_0\approx 7R_E$
and $\Theta=\arctan(|B_y|/B_z)$, and was calculated from the WIND spacecraft
key parameter database \citep{lepping,ogilvie}.  The indices and $\epsilon$
are from different time intervals and here we assume statistical stability
over these long time intervals.
\begin{figure}
\epsfsize=0.45\textwidth \centerline{
\leavevmode\epsffile{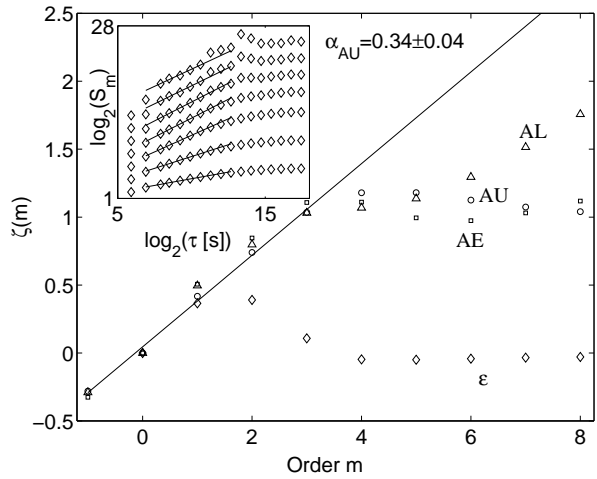}} \caption{Dependence of the scaling exponent
$\zeta(m)$ of the raw GSF on moment order $m$. Inset shows the GSF $S_m$
versus time lag $\tau$ for $AU$.}
\label{fig2}
\end{figure}
\begin{figure}
\epsfsize=0.45\textwidth \centerline{
\leavevmode\epsffile{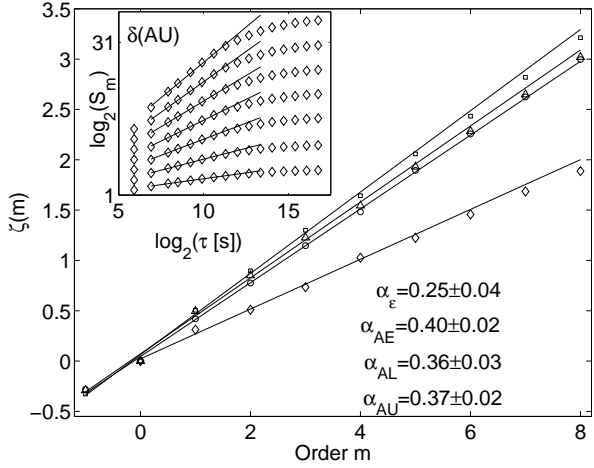}} \caption{Dependence of the
scaling exponent of the conditioned GSF on moment order. Inset
shows the conditioned GSF $S^c_m$ versus time lag $\tau$ for $AU$.}
\label{fig3}
\end{figure}

\section{Generalized Structure Functions}
Generalized structure functions (GSF), or generalized variograms,
can be defined in terms of an average over time of a differenced
variable $\delta x(t,\tau)=x(t+\tau)-x(t)$ as
$S_m(\tau)=<|\delta x(t,\tau)|^m>$ \citep{rodriguez}. If $\delta x$ exhibits
scaling with respect to $\tau$, then $S_m\propto\tau^{\zeta(m)}$.
A log-log plot of $S_m$ versus $\tau$ should then reveal a straight line for
each $m$ with gradients $\zeta(m)$. If $\zeta(m)=\alpha m$ ($\alpha$ constant)
then the time series is self-similar with single scaling exponent $\alpha$.

In order to compare the scaling properties of the non-contemporaneous
$\epsilon$ and $AE$ indices time series, we select a $4$-year subinterval
$1984-1987$ from the $AE$ indices at the same phase in the solar cycle as the
$\epsilon$ data. Figure \ref{fig1} shows the second order GSFs as measured by
the standard deviations $\sigma(\tau) = [S_2(\tau)]^{1/2}$ of the fluctuation
$\delta x(t,\tau)$.  A scaling region is apparent between $2^7$ and $2^{12}$~s
where $\sigma(\tau)\propto\tau^H$, where $H$ is the Hurst exponent
[$\zeta(2)/2$]. The $R^2$ goodness of fit analysis was performed to select the
optimal power law region and gradient and results are summarized in Table
\ref{tab1}. The upper limits of the scale regions $\tau_{max}$ are in good
agreement with values reported previously \citep{consolini98,takalo93,takalo98}.

Any such single estimate of the $H$, whilst establishing the region of $\tau$
over which there is scaling, does not fully characterize the properties of the
time series. For example, a fractional Brownian motion (fBm) can be constructed 
to share the same $H$ value as $AE$, but the fBm series has Gaussian distributed
increments $\delta x$ by definition \citep{mandelbrot} whereas those of $AE$
are non-Gaussian \citep{consolini98,hnat02}. As discussed by \cite{mandelbrot}
the similar values arise because $H$ aggregates {\em two} sources of scaling
in monofractal random walks: persistence (the ``Joseph" effect) and heavy
tails in the increments (the ``Noah" effect). In the above example the
anomalous value of $H$ for fBm comes just from the Joseph effect, whilst
for $AE$ the Noah effect must be at work. Furthermore, estimating $H$ by only
one method may not distinguish a fractal time series from a discontinuous
one \citep{watkins,katsev}. We thus turn next to the higher order $m$ values
of $\zeta(m)$.

Figure \ref{fig2} shows scaling exponents $\zeta(m)$ derived from raw GSFs with
$m$ varying between $-1$ and $8$ for the $\delta\epsilon$ and $AE$ indices
fluctuations. These suggest the departure of higher orders from
self-similarity, i.e., $\zeta(m)$ departs from a straight line. The inset of
this figure shows the origin of these $\zeta(m)$ values for $\delta AU$ and
$m=1,...,7$. Only the first four orders exhibit clear linear behavior expected
in the scaling region. For higher orders, the value of $\zeta$ very strongly
depends on the assumed extent of the scaling region to which one fits a
straight line. In principle, $\zeta(m)$ can be obtained for any $m$.
However, errors do not contribute uniformly over $m$, for example,
the largest fluctuations that affect large $m$, are statistically
poorly resolved, whereas the smallest fluctuations
($\delta x\rightarrow 0$) are dominated by instrument thresholds. For the
latter reason we will exclude $m=-1$ for $\delta \epsilon$ as
$\delta\epsilon\rightarrow 0$ is not well determined through its definition.

Conditioned GSFs quantify the impact of intermittency on fluctuations of
different sizes by imposing a threshold $A$ on the event size \citep{kovacs}.
Here, this threshold will be based on the standard deviation of the
differenced time series for a given $\tau$, $A(\tau)=10\sigma(\tau)$. This
procedure allows us to exclude rare extreme fluctuations with large statistical
errors which, for large $m$, could lead to a spurious departure from
self-similar behavior.  Alternatively, conditioning with different thresholds
estimates a maximum size for the fluctuations for which self-similarity is
still valid.

Following conditioning, log-log plots of $S^c_m(\tau)$ show good correspondence
with straight line fits, shown for $\delta AU$ in the inset of figure
\ref{fig3}. This power law dependence holds between times already obtained
from the $R^2$ analysis performed for $\sigma(\tau)$. The main plot then shows
$\zeta(m)$ obtained from the conditioned $S^c_m(\tau)$. All lines in the figure
were fitted for moments between $-1$ ($0$ for $\epsilon$) and $6$ and then
extended to the entire range of data. Scaling exponents obtained from this
technique were unchanged for thresholds $A(\tau)$ between $6\sigma$ and
$12\sigma$.

Firstly, our analysis suggests that the statistics of the fluctuations for all
four quantities are self-similar for times between $2$ and $\sim100$ minutes
and fluctuations of size $\delta x\leq 10\sigma(\tau)$. Secondly, the
scaling exponent $\alpha$ in $\zeta(m)=\alpha m$ that characterize this
self-similar behavior, are identical within errors for fluctuations in the AE
indices but different to that in $\epsilon$ at the $1\sigma$ level.

\section{Rescaling of Fluctuation PDFs}
Scaling of the GSFs can be related to scaling properties of the fluctuation
PDFs \citep{hnat02,hnat03a} using the generic, model-independent rescaling
method (e.g. \citep{stanley,hnat03b}) based on the rescaling of the PDFs
$P(\delta x,\tau)$ of $\delta x(t,\tau)$ on different time scales $\tau$.
If a time series exhibits statistical self-similarity, a single argument
representation of the PDF can be found that is given by
$P(\delta x,\tau)=\tau^{-\alpha} P_s(\delta x \tau^{-\alpha})$,
where $\alpha$ is the rescaling exponent. We now express $S_m$ using the
fluctuations' PDF, $P(\delta x,\tau)$ as follows:
\begin{equation}
S_m(\tau)=\int_{-\infty}^{\infty}|\delta x|^mP(\delta x,\tau)d(\delta x).
\label{strpdf1}
\end{equation}
Expressing the integral in (\ref{strpdf1}) in terms of rescaled variables
$P_s$ and $\delta x_s=\delta x \tau^{-\alpha}$ shows that the scaling exponent
$\zeta(m)$ is a linear function of $m$, $\zeta(m)=m\alpha$, for a
statistically self-similar process, as suggested here by figure \ref{fig3}.

The exponent $\alpha$ is ideally obtained from the scaling of the peaks of the
PDF $P(0,\tau)$. However, the finite accuracy of the measurement may discretize
the amplitude leading to errors in the peak values.
Table \ref{tab1} gives all scaling exponents, obtained by different methods.
These yield consistent values of $\alpha$, to within the errors. We will use
$\alpha$ from the scaling of $\sigma(\tau)$ versus $\tau$. If the fluctuations
are statistically self-similar, as suggested by our GSF analysis, then the
unscaled PDFs $P(\delta x,\tau)$ should collapse onto a single curve
$P_s(\delta x_s)$. We applied PDF rescaling to the fluctuation PDFs
of all quantities and obtained satisfactory collapse of the curves within the scaling regions. The $\chi^2$ test applied to all quantities revealed that,
for the scaling regions given above, the collapsed curves lie within $5-7\%$
error band.
\begin{table}[b]
\begin{center}
\begin{tabular}{|p{1.1cm}|p{1.85cm}|p{1.55cm}|p{1.6cm}|p{1.4cm}|}
\hline
Quantity&$\alpha$ from $P(0,\tau)$&$\alpha$ from $\sigma(\tau)$&$\alpha$ from GSF&$\tau_{max} [min]$\\
\hline
$\epsilon$&$-----$&$0.31\pm0.04$&$0.25\pm0.04$&$\sim100$\\
\hline
AE&$-0.47\pm0.03$&$0.46\pm0.02$&$0.40\pm0.02$&$\sim100$\\
\hline
AU&$-0.46\pm0.03$&$0.41\pm0.02$&$0.37\pm0.02$&$\sim60$\\
\hline
AL&$-0.45\pm0.03$&$0.44\pm0.02$&$0.36\pm0.03$&$\sim100$\\
\hline
\end{tabular}
\caption{Scaling indices derived from $P(0,\tau)$, $\sigma(\tau)$
and GSF power laws.}
\end{center}
\label{tab1}
\end{table}

\begin{figure}
\epsfsize=0.45\textwidth \centerline{
\leavevmode\epsffile{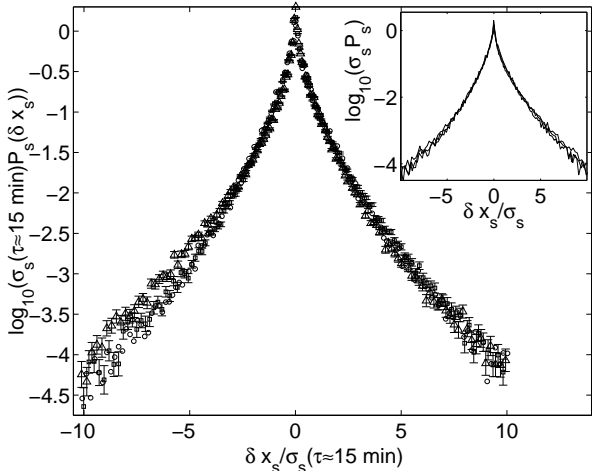}} \caption{Direct comparison
between the fluctuation PDFs for $AE$($\square$), $AU$($\circ$) and
$AL$($\triangle$), again at $\tau=15$ minutes. Inset shows overlaid
PDFs of $AU$, $AE$ and $-AL$ fluctuations. Error bars as in Figure 1.}
\label{fig4}
\end{figure}

Figure \ref{fig4} shows the re-scaled fluctuation PDFs for the indices alone
for $\tau\approx 15$ min. The $\delta x$ variable has been normalized to the
rescaled standard deviation $\sigma_s(\tau\approx 15 min.)$ of $P_s$ in each
case to facilitate this comparison. The inset of this figure shows the
comparison for $AU$, $AE$ and $-AL$ fluctuations and these PDFs are nearly
identical. These results are consistent with conclusions of the GSF analysis
at the $1\sigma$ level.
\begin{figure}
\epsfsize=0.45\textwidth \centerline{
\leavevmode\epsffile{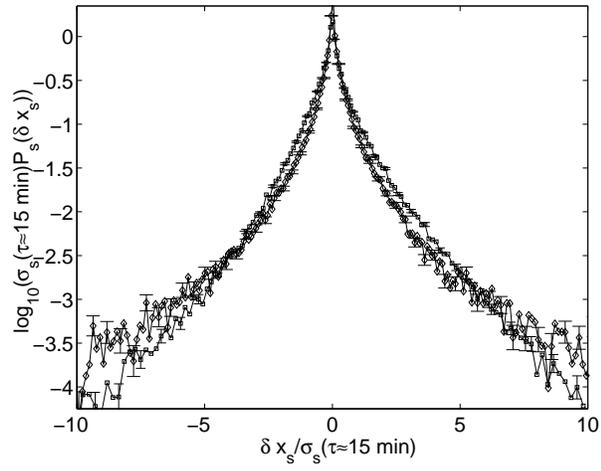}} \caption{Direct comparison,
for the particular choice $\tau=15$ minutes, of the fluctuation
PDFs for $\epsilon$ ($\diamond$) and $AE$ index ($\Box$). Error bars
as in Figure 1.}
\label{fig5}
\end{figure}

Figure \ref{fig5} shows the normalized PDFs $P_s(\delta x_s)$ for
$\delta x=\delta \epsilon$, $\delta AE$ and $\tau \approx 15$ min overlaid on
a single plot. We can clearly distinguish between the PDFs of the
$\delta \epsilon$ and $AE$ indices' fluctuations. We obtain the same result
repeating this comparison for several values of $\tau$, within the scaling
range $\tau_{max}$. We have also verified that the functional form of the
PDF are insensitive to the solar cycle within errors. The use of a larger,
$10$ year data set for the indices has reduced statistical scatter and
expanded the dynamic range of the considered fluctuations as compared to
the analysis given in \citep{hnat02,hnat03a}, and would lead us to draw the
opposite conclusion, that on time scales less than $\approx1$ hour the $AE$
index amplitude fluctuations are not driven linearly by those of the solar wind.
We would also conclude that the difference seen at the $1\sigma$ level in the
scaling of the $\epsilon$ and the indices is significant, even though they
agree at the $2\sigma$ level \citep{freeman}.

\section{Summary}
In this paper we have addressed an open question of the possible connection
between the scaling properties of fluctuations in the solar wind driver and
those observed in global measures of magnetospheric dynamics. We applied two
statistical methods, generalized structure functions and PDF rescaling, to
study the scaling of fluctuations in the $\epsilon$ parameter and the
magnetospheric indices $AU$, $AL$ and $AE$. We find that, statistically,
fluctuations in all four quantities are approximately self-similar when their
size is limited to $\sim10\sigma$. This self-similarity extends to $\sim 1-1.5$
hours. The scaling exponents of the $AE$ indices are close to each other and
are appreciably different to that of the $\epsilon$ parameter.

The fluctuation PDFs of the $AE$ indices, unlike that of $\delta \epsilon$, are
asymmetric. Direct comparison of the PDFs for the fluctuations in the $AU$,
$AE$ and $-AL$ index indicates that they are nearly identical.
Whilst the low frequency behavior of the solar wind and the indices may be
well correlated \citep{tsurutani}, here we have concluded that, on time scales
smaller than $1$ hour the properties of the fluctuations in the solar wind and
the indices differ in {\em both} amplitude and persistence. If the underlying
physical origin of the auroral scaling is turbulence, then different scaling
behavior implies a different type of turbulence, i.e., different
dimensionality/topology or different relevant physics \citep{frisch}.
If the underlying physics is SOC or similar \citep{chang03} then similar
conclusions would still be drawn (c.f. \citep{uritsky}). However, at this
point we also can not rule out the possibility that the way in which the
indices are constructed ``burns" information still present in the
magnetometer data about the solar wind scaling, here possibly by changing
either or both of the degree of persistence (power spectral slope) and the 
heavy-tailed property (see \citep{edwards} for a related preliminary
investigation).

\section{Acknowledgment}
SCC and BH acknowledge the PPARC and GR the Leverhulme Trust.
We thank R.~P.~Lepping and K.~Ogilvie for provision of data from the
NASA WIND spacecraft and the World Data Center C2, Kyoto for geomagnetic
indices.
%%%%%%%%%%%%%%%%%%%%%%%%%%%%%%%%%%%%%%%%%%%%%%%%%%%%%%%%%%%%%%%%%%%%%%

\end{article}
\end{document}